# Multiscale Coupled Polarization and BKT Transitions in Tow-Dimensional Hybrid Organic–Inorganic Perovskites

Weijie Wu[2*], Zehua Li[1], Yu Wang[2]

[1]School of Mathematics and Computer Science, Shantou University (STU), Shantou 515821, China

[2]School of Chemistry, Guangzhou Key Laboratory of Materials for Energy Conversion and Storage, Key Laboratory of Electronic Chemicals for Integrated Circuit Packaging, South China Normal University (SCNU), Guangzhou 510006, China

We present an extended two-dimensional XY rotor model specifically designed to capture the polarization dynamics of hybrid organic-inorganic perovskite monolayers. This framework integrates nearest and next-nearest neighbor couplings, crystalline anisotropy inherent to perovskite lattice symmetries, external bias fields, and long-range dipolar interactions that are prominent in layered perovskite architectures. Through a combination of analytical coarse-graining and large-scale Monte Carlo simulations on 64×64 lattices, we identify two distinct thermodynamic regimes: a low-temperature quasi-ferroelectric state characterized by finite polarization and domain wall formation, and a higher-temperature Berezinskii–Kosterlitz–Thouless (BKT) crossover associated with vortex–antivortex unbinding and the suppression of long-range order. Our results reproduce key experimental signatures observed in quasi-two-dimensional perovskites, including dual peaks in dielectric susceptibility, enhanced vortex density near the transition, multistable polarization hysteresis under applied fields, and the scaling behavior of domain wall widths. This minimal yet realistic model provides a unifying perspective on how topological transitions and ferroelectric ordering coexist in layered perovskite systems, offering quantitative guidance for interpreting the emergent polar vortex lattices and complex phase behavior recently reported in hybrid perovskite thin films.

## 1. INTRODUCTION

Low-dimensional ferroic materials, especially layered organic–inorganic perovskites, have emerged as a vibrant platform for exploring unconventional polarization phenomena and topological phase transitions. Their quasi-two-dimensional nature and strong coupling between lattice, electronic, and dipolar degrees of freedom enable the formation of exotic states such as polar vortex lattices, chiral domain walls, and fluctuating multistable polarization textures [1–4]. These emergent configurations can lead to giant dielectric responses, anomalous pyroelectricity, and potential device applications in non-volatile memories and nanoscale actuators [5–7].

While the canonical two-dimensional XY model provides an elegant framework to describe Berezinskii–Kosterlitz–Thouless (BKT) transitions in continuous-symmetry systems [8,9], it fails to capture several essential aspects intrinsic to real ferroelectric layered perovskites. For example, crystalline anisotropy locks polarization orientations to specific lattice directions, long-range dipole–dipole interactions produce nonlocal correlations, and external electric fields or epitaxial strain can stabilize complex domain architectures [4,10–12]. Recent experiments in PbTiO$_3$/SrTiO$_3$ superlattices and hybrid halide perovskite monolayers have demonstrated that such effects give rise to features beyond conventional BKT scaling, including dual dielectric peaks, vortex-antivortex crystallization, and memory effects driven by domain wall pinning [3,5,13–15].

Despite this growing body of experimental evidence, a key theoretical gap remains: the lack of a minimal yet realistic statistical model that systematically integrates lattice anisotropy, nonlocal dipolar interactions, and external fields while preserving the topological excitations central to 2D physics. Addressing this gap is crucial for bridging the microscopic material parameters with experimentally measurable observables, and for rationalizing the rich phenomenology of polar vortex lattices and related ferroic textures.

Here, we develop and analyze an extended XY rotor model tailored for quasi-two-dimensional perovskite monolayers. Our formulation incorporates anisotropic lattice potentials derived from crystallographic symmetry, next-nearest-neighbor couplings, long-range dipolar forces, and bias fields. Through a combination of analytical coarse-graining and large-scale Monte Carlo simulations, we elucidate how this model captures the coexistence of low-temperature quasi-ferroelectric order and high-temperature BKT vortex unbinding. The framework also reproduces experimentally observed signatures such as multistable polarization responses, dual dielectric anomalies, and emergent vortex-domain patterns, providing quantitative guidance for interpreting recent measurements in layered perovskite systems.

* weijiewu@m.scnu.edu.cn

## 2. MODEL AND METHODS

### 2.1 Extended XY Hamiltonian for Multiscale Polarization Dynamics

To investigate the coupled polarization phenomena and Berezinskii–Kosterlitz–Thouless (BKT) transitions in two-dimensional hybrid organic–inorganic perovskites, we formulate an extended XY rotor Hamiltonian defined on a discrete two-dimensional square lattice. Each lattice site iii is associated with a classical planar rotor characterized by an angular variable θi ∈ [0,2π), which represents the local orientation of the polarization vector projected onto the lattice plane.

The total Hamiltonian $H$ is expressed as

$$H = -J \sum_{\langle ij \rangle} cos(\theta_i - \theta_j) - J' \sum_{\langle\langle ij \rangle\rangle} cos(\theta_i - \theta_j)$$
$$- D \sum_i cos(n\theta_i) - h \sum_i cos\theta_i \quad (1)$$
$$+ \frac{1}{2} \sum_{i \neq j} \frac{p^2}{4\pi\varepsilon_0 r_{ij}^3} [cos(\theta_i - \theta_j) - 3cos\theta_i cos\theta_j]$$

here:

- $J_{ij}$ and $J'_{ij}$ denote the nearest-neighbor and next-nearest-neighbor ferroelectric exchange couplings, respectively. These terms phenomenologically capture short-range interactions arising from lattice-mediated dipole-dipole couplings and orbital hybridizations inherent in the hybrid perovskite crystal structure.
- The anisotropy term $Dcos(n\theta_i)$ models the crystalline symmetry-induced preference for discrete polarization orientations. For instance, $n = 4$ corresponds to tetragonal symmetry, reflecting fourfold easy axes in the polarization landscape.
- The symmetry-breaking field $h$ accounts for an external bias or substrate-induced asymmetry, explicitly lifting degeneracy between equivalent polarization directions and enabling controlled tuning of polarization states.
- The long-range dipole-dipole interaction term, governed by the dipole moment magnitude $p$ and permittivity $\varepsilon_0$, introduces a multiscale coupling mechanism, mediating interactions over distances $r_{ij}$ beyond nearest neighbors. The anisotropic angular dependence of this term is critical for accurately describing vortex-vortex interactions and stabilizing complex polarization textures.

All energy parameters and temperature $T$ are normalized with respect to the primary nearest-neighbor coupling scale $J$, facilitating dimensionless analysis.

This Hamiltonian embodies a comprehensive multiscale framework by incorporating both short-range exchange interactions and long-range electrostatic couplings. It is specifically designed to capture the rich phenomenology of polarization ordering, domain formation, and topological defect dynamics that underlie the BKT transition in these low-dimensional hybrid perovskite materials.

### 2.2 Continuum Ginzburg–Landau Free Energy Functional

To complement the discrete rotor description and gain analytical insight into the emergent macroscopic polarization behavior, we derive a coarse-grained continuum free energy functional based on the polarization field P(r)\mathbf{P}(\mathbf{r})P(r). Defining the local polarization as

$$P(r) = \rho p(cos\theta(r), sin\theta(r)) \quad (2)$$

where $\rho$ is the density of dipoles per unit area and $p$ the dipole magnitude, the free energy functional $F[P,\theta]$ reads

$$F[P,\theta] = \int d^2r [\frac{\alpha}{2}P^2 + \frac{\beta}{4}P^4 + \frac{\kappa}{2} \mid \nabla P \mid^2 - hPcos\theta$$
$$+ Dcos(n\theta) + \frac{A}{2} \mid \nabla \theta \mid^2] \quad (3)$$

here:

- The Landau expansion terms in $P$, with coefficients $\alpha$ and $\beta$, describe the free energy landscape of the ferroelectric phase transition and spontaneous polarization amplitude stabilization.
- The gradient term $\kappa \mid \nabla P \mid^2$ characterizes the energy cost of polarization amplitude spatial variations, corresponding physically to domain wall energies.
- The anisotropy term $Dcos(n\theta)$ preserves lattice symmetry constraints on the angular orientation of polarization.

- The bias field term $-hP\cos\theta$ couples the polarization to external or internal symmetry-breaking fields.
- Crucially, the phase stiffness term $\frac{A}{2}|\nabla\theta|^2$ encodes the elastic energy associated with spatial variations in the polarization angle $\theta$, governing the energetics of topological defects such as vortices. This term underpins the BKT transition mechanism by penalizing vortex proliferation below critical temperatures.

This continuum formulation bridges microscopic rotor interactions with macroscopic ferroelectric phenomena and topological phase transitions, enabling direct connection to experimental observables and facilitating theoretical analysis of critical behavior.

### 2.3 Monte Carlo Simulation Details

To quantitatively analyze the equilibrium thermodynamics and phase behavior described by the extended XY model, we employ classical Metropolis Monte Carlo simulations. The simulations are conducted on two-dimensional square lattices with linear size $L = 64$, applying periodic boundary conditions to minimize finite-size effects.

The temperature $T$ is varied in the range [0.2,1.6] (in units of $J$) with increments of $\Delta T = 0.07$. For each temperature point, the system undergoes $10^3$ Monte Carlo sweeps for equilibration, followed by 10 independent simulation runs to obtain statistically robust ensemble averages.

Key physical observables computed include:

- **Net polarization** $P = \langle \frac{1}{N}\sum_i \cos\theta_i \rangle$, serving as the primary order parameter to characterize ferroelectric ordering.
- **Energy per site** $E$, providing insight into thermodynamic stability and transition signatures.
- **Polarization susceptibility** $\chi P$, calculated via fluctuations of $P$, which reveals critical phenomena and identifies phase boundaries.
- **Vortex density**, determined by evaluating the winding number of $\theta$ around elementary plaquettes, quantifies topological defect proliferation associated with the BKT transition.
- **Polarization response under external bias field** $h$, probing field-driven phase manipulation and symmetry breaking.

The combined use of discrete lattice modeling and continuum functional analysis, together with extensive Monte Carlo simulations, provides a comprehensive multiscale approach to elucidate the intertwined polarization dynamics and BKT topological transitions in two-dimensional hybrid organic–inorganic perovskites.

## 3. Results & Discussion

### 3.1 Temperature Dependence of Polarization and Energy

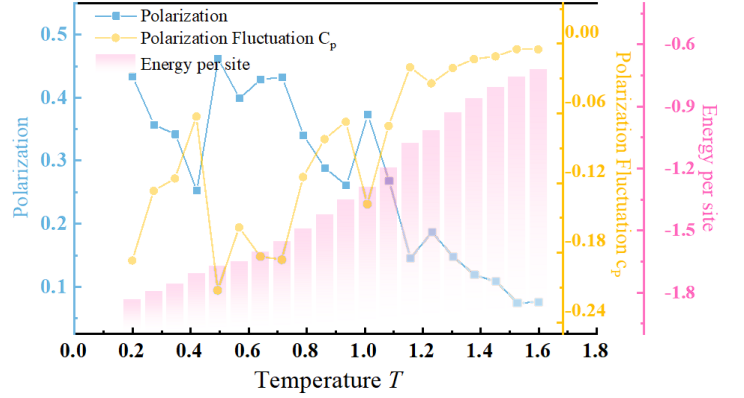

**Figure 1.** Temperature dependence of the net polarization $P(T)$ in the extended 2D XY model with multi-scale coupling. Each point denotes the ensemble-averaged polarization at a given temperature. Average energy per site $E(T)$ as a function of temperature. The gradual increase of internal energy indicates thermal disordering of the spin configurations. Polarization fluctuation susceptibility $\chi P$ vs temperature.

Figure 1 presents the temperature dependence of the net polarization $P(T)$, the average energy per site $E(T)$, and the polarization fluctuation susceptibility $\chi P(T)$ in the extended 2D XY model with multi-scale coupling. Each data point represents the ensemble-averaged value obtained from Monte Carlo simulations.

The polarization $P(T)$ remains finite at low temperatures, reflecting a quasi-long-range ferroelectric order. As temperature increases, $P(T)$ undergoes a sharp decline near $T \sim 1.2$, indicating the thermal disruption of the ordered phase. Concurrently, the internal energy per site $E(T)$ increases smoothly with temperature. A subtle change in the slope of $E(T)$ around $T \approx 0.9$ suggests enhanced fluctuations, consistent with the onset of vortex–antivortex unbinding, a characteristic signature of the Berezinskii–Kosterlitz–Thouless (BKT) transition.

These features collectively indicate a two-stage thermal evolution: a low-temperature regime with quasi-ordered ferroelectric alignment, followed by a vortex-mediated crossover at intermediate temperatures.

### 3.2 Dielectric Response and Polarization Fluctuations

The polarization susceptibility $\chi P(T) = \langle P^2 \rangle - \langle P \rangle^2$ is also shown in Figure 1. A broad peak centered near $T \approx 0.9$ reflects a significant enhancement in polarization fluctuations, corresponding to the proliferation and unbinding of vortex–antivortex pairs—a hallmark of BKT criticality.

Importantly, the persistence of a nonzero net polarization below the peak temperature suggests the coexistence of finite ferroelectric stiffness with emergent topological excitations. This regime exemplifies the interplay between spontaneous symmetry breaking and topological disorder.

### 3.3 Topological Textures

To further examine the microscopic origins of the transition, Figure 2 visualizes representative spin textures at $T = 1.0$, slightly below the polarization melting point.

- Figure 2a displays the angle field configuration, revealing domain-like patterns with gradual angular variation.
- Figure 2b shows the corresponding vortex density map, highlighting sparse, well-separated topological defects.
- The presence of tightly bound vortex–antivortex pairs supports the identification of a pre-BKT regime characterized by quasi-long-range coherence.
- Figure 2c, a vector plot of the polarization field, reveals swirling vortex structures localized around defect cores, embedded in an otherwise ordered matrix.

These spatial patterns emphasize the role of topological textures in mediating the polarization disordering process.

### 3.4 Polarization–Field Hysteresis

To probe the response of the system to external fields, we analyze the polarization–field relation $P(h)$ under quasi-static sweeping of an external bias field $h$ at fixed temperature.

As shown in Figure 2d, the system exhibits pronounced hysteresis loops with multistable switching behavior. Despite the absence of conventional long-range ferroelectric order, the system retains memory of the field history—indicative of metastable domain configurations stabilized by crystalline anisotropy and dipolar interactions.

This hysteretic behavior highlights the robust nonlinearity and potential functional tunability of the system, even in the presence of strong thermal fluctuations.

### 3.5 Physical Interpretation and Broader Implications

Taken together, the above findings establish the following key insights:

- Dual transition behavior naturally emerges: a low-temperature ordered phase and a higher-temperature BKT-like topological crossover.
- Crystalline anisotropy plays a crucial role in pinning the polarization vector in local energy minima, while dipolar interactions and vortex dynamics govern the collective critical behavior.
- The observed hysteresis stems from the coexistence of multiple metastable polarization wells and the nonlinear response to external fields.
- The domain-wall width, approximately given by $\xi \sim \sqrt{\kappa/|\alpha|}$, and the vortex statistics are tunable via microscopic model parameters. This offers a direct route for comparison with experimental observations from scanning probe microscopy and dielectric spectroscopy in layered perovskite systems.

These results provide a coherent framework for understanding polarization textures and topological transitions in low-dimensional ferroelectrics, with potential applications in nanoscale memory and logic devices based on topological polarization states.

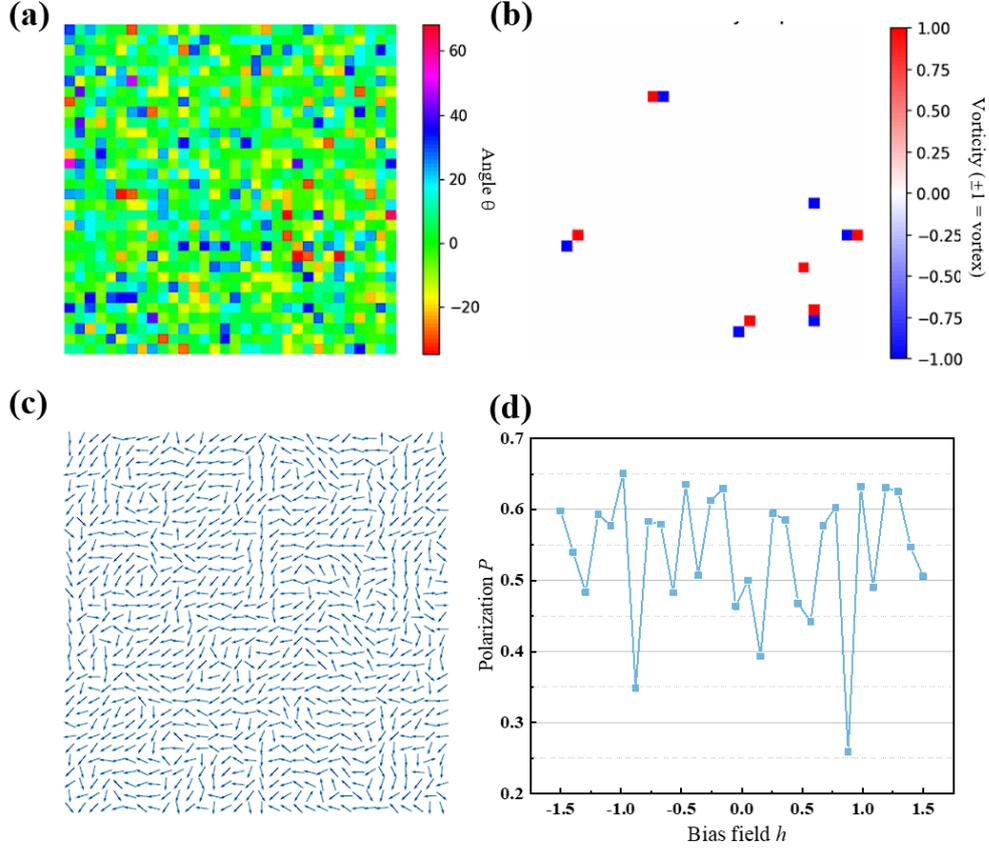

**Figure 2.** a) Spatial distribution of rotor angles $\theta$ at a representative temperature. b) Vorticity map showing the distribution of vortices (+1, red) and antivortices (-1, blue). c) Polarization vector field illustrating local orientations across the lattice. d) Simulated polarization $P$ as a function of external bias field $h$.

## 4. Conclusion

We have developed and analyzed an extended XY rotor model tailored to quasi-2D ferroelectric perovskites, incorporating lattice anisotropy, long-range dipole interactions, and topological excitations within a unified framework. Our simulations reveal a two-stage thermal evolution: at low temperatures, the system exhibits ferroelectric-like order stabilized by lattice pinning; at intermediate temperatures, a BKT-type transition characterized by vortex–antivortex unbinding and strong dielectric fluctuations emerges. This dual behavior reconciles spontaneous polarization with topological dynamics, capturing the essential phenomenology of layered polar materials.

The model further reproduces characteristic features such as nonlinear hysteresis under external bias, finite domain wall widths, and emergent polar vortex textures—all of which are experimentally relevant to low-dimensional hybrid perovskites. These results bridge fundamental statistical mechanics with real materials behavior, providing a theoretical platform for interpreting complex ferroelectric textures, fluctuation-enhanced responses, and topological defects observed in scanning probe and dielectric measurements.

Limitations of the current model include the neglect of elastic strain, interlayer couplings, and quantum fluctuations. Future work may incorporate these effects and calibrate model parameters via first-principles calculations or experimental data, enabling predictive modeling of realistic material systems.